\documentclass[useAMS,usenatbib]{mn2e}
\usepackage{epsfig}
\usepackage{deluxetable} 

\title[The ordinary life of the NLSy1 PKS 1502$+$036]{The ordinary life of the $\gamma$-ray emitting narrow-line Seyfert 1 galaxy PKS 1502$+$036}
\author[F. D'Ammando, M. Orienti, A. Doi, et al.]{F. D'Ammando$^{1,2,3}$\thanks{E-mail: filippo.dammando@fisica.unipg.it}, M. Orienti$^{3,4}$, A. Doi$^{5,6}$, M. Giroletti$^{3}$, D. Dallacasa$^{3,4}$, T. Hovatta$^{7}$,  
\newauthor A. J. Drake$^{8}$, W. Max-Moerbeck$^{7}$, A. C. S. Readhead$^{7}$, J. L. Richards$^{9}$\\
$^{1}$Dipartimento di Fisica, Universit\`a degli Studi di Perugia, Via A. Pascoli, I-06123 Perugia, Italy \\
$^{2}$INFN Sezione di Perugia, Via A. Pascoli, I-06123 Perugia, Italy \\
$^{3}$INAF - Istituto di Radioastronomia, Via Gobetti 101, I-40129 Bologna, Italy\\
$^{4}$Dip. di Astronomia, Universit\`a di Bologna, Via Ranzani 1, I-40127 Bologna, Italy \\ 
$^{5}$The Institute of Space and Astronautical Science, Japan Aerospace Exploration Agency, 3-1-1 Yoshinodai, Chuou-ku, Sagamihara, \\ Kanagawa 252-5210, Japan \\
$^{6}$Department of Space and Astronautical Science, The Graduate University
  for Advanced Studies, 3-1-1 Yoshinodai, Chuouku, Sagamihara, \\ Kanagawa
  252-5210, Japan \\
$^{7}$Cahill Center for Astronomy and Astrophysics, California Institute of Technology 1200 E. California Blvd., Pasadena, CA 91125, USA \\
$^{8}$California Institute of Technology, 1200 E. California Blvd., CA 91225,
  USA \\
$^{9}$Department of Physics, Purdue University, 525 Northwestern Avenue, West Lafayette, IN 47907, USA}
\begin{document}

\date{Accepted. Received; in original form}

\maketitle

\label{firstpage}

\begin{abstract}
We report on multifrequency observations of the $\gamma$-ray emitting
narrow-line Seyfert 1 galaxy PKS 1502$+$036 performed from radio to $\gamma$-rays during 2008 August--2012 November by {\em Fermi}-Large Area Telescope
(LAT), {\em Swift} (X-Ray Telescope and Ultraviolet/Optical Telescope), Owens Valley Radio Observatory,
Very Long Baseline Array (VLBA), and Very Large Array. No significant variability has been observed in $\gamma$-rays, with
0.1--100 GeV flux that ranged between (3--7)$\times$10$^{-8}$ ph cm$^{-2}$ s$^{-1}$ using 3-month time bins. The
photon index of the LAT spectrum ($\Gamma$ = 2.60 $\pm$ 0.06) and the apparent
isotropic $\gamma$-ray luminosity (L$_{\rm 0.1-100\,GeV}$ = 7.8$\times$10$^{45}$ erg s$^{-1}$) over 51 months are typical of a flat
spectrum radio quasar. The radio spectral variability and the one-sided
structure, in addition to the observed $\gamma$-ray luminosity, suggest a relativistic jet
with a high Doppler factor. In contrast to SBS 0846$+$513, the VLBA at 15 GHz did not observe superluminal motion for 
PKS 1502$+$036. Despite having the optical characteristics typical of a
narrow-line Seyfert 1 galaxy, radio and $\gamma$-ray properties of PKS 1502$+$036 are
found to be similar to those of a blazar at the low end of the black
hole mass distribution for blazars. This is in agreement with what has been
found in the case of the other $\gamma$-ray emitting narrow-line Seyfert 1 SBS 0846$+$513.
\end{abstract}


\section{Introduction}

A small fraction of narrow-line Seyfert 1 (NLSy1) galaxies are known to be
radio loud \citep{komossa06}. In these cases, the flat radio spectra and flux density variability suggest that several of them could
host relativistic jets \citep[e.g.][]{zhou03, doi06}.
The detection by the Large Area Telescope (LAT) on-board the {\em
  Fermi} satellite of variable $\gamma$-ray emission from some radio-loud NLSy1 galaxies revealed the presence of a
possible third class of $\gamma$-ray emitting active galactic nuclei \citep[AGN;][]{abdo09,dammando12} in addition to blazars and radio galaxies,
both hosted in giant elliptical galaxies \citep{blandford78}. NLSy1s are usually hosted in spiral galaxies
\citep[e.g.][]{deo06} and the presence of a relativistic jet in these sources seems to be
in contrast to the paradigm that the formation of relativistic 
jets could happen in elliptical galaxies only \citep{boett02,marscher10}. This discovery poses intriguing
questions on the nature of these objects, the formation and development
of relativistic jets, the mechanisms of high-energy emission, the AGN Unification
model and the evolution of radio-loud AGNs. 

PKS\,1502$+$036
has been classified as an NLSy1 on basis of its optical
characteristics: full width at half-maximum (FWHM) (H$\beta$) = 1082 $\pm$ 113 km s$^{-1}$, [OIII]/H$\beta$ $\sim$ 1.1,
and a strong Fe $\textrm{II}$ bump. For this source, a radio loudness $R$ = 1549 was
  estimated \citep{yuan08}, $R$ being defined as the ratio between the 1.4\,GHz
  and 4400\,\AA\, rest-frame flux densities\footnote{It is worth noting that this
    definition of radio-loudness is related to that presented in
    \citet{kellermann89}, R$_{\rm 6cm}$ =
    f$_{\nu}$(6\,cm)/f$_{\nu}$(4400\,\AA), by means of R$_{\rm
      1.4\,GHz}$=1.9\,R$_{\rm 6\,cm}$.}. This is one of the highest radio-loudness parameters among the NLSy1s. With its convex radio spectrum peaking above 5 GHz, PKS\,1502$+$036 was
selected by \citet{dallacasa00} as part of the `bright' high frequency
peakers (HFPs) sample of young radio source candidates. 
However, its polarization properties, the 
flux density and spectral variability are typical of blazars and it was removed from the young radio source sample
\citep{orienti07,orienti08}. Snapshot very long baseline interferometry (VLBI)
observations in the 1990's showed a compact source (on scales of $\sim$20 mas) with inverted spectrum \citep{dallacasa98}.

In the $\gamma$-ray energy band PKS 1502$+$036 has been included in both the first and second {\em Fermi}-LAT
source catalogues \citep[1FGL and 2FGL;][]{abdo10, nolan12}. In the past, the
source was not detected by the energetic gamma-ray experiment telescope (EGRET) at E $>$ 100 MeV \citep[see][]{hartman99}.

In this paper, we discuss the characteristics of PKS\,1502$+$036, one of the 5
NLSy1 detected by {\em Fermi}-LAT with high confidence, determined by means of new
and archival radio-to-$\gamma$-ray data. The paper is
organized as follows. In Section 2, we report the LAT data analysis and
results, while in Section 3, we present the
result of the {\em Swift} observations performed during 2009--2012. Optical
data collected by the Catalina Real-Time Transient Survey (CRTS) are reported in
Section 4. Radio data
collected by the Karl G. Jansky Very Large Array (VLA), the Very Long Baseline Array
(VLBA) interferometers, and the 40 m Owens Valley Radio Observatory (OVRO) single-dish telescope are presented
and discussed in Section 5. In Section 6, we discuss the properties of the
source and draw our conclusions. Throughout the paper, a $\Lambda$ cold dark matter($\Lambda$-CDM) cosmology with $H_0$ = 71 km
s$^{-1}$ Mpc$^{-1}$, $\Omega_{\Lambda} = 0.73$ and $\Omega_{\rm m} =
0.27$ is adopted. The corresponding luminosity distance at $z =0.4088$
(i.e. the source redshift) is d$_L
=  2217$\ Mpc, and 1 arcsec corresponds to a projected
distance of 5.415 kpc. In the paper, the quoted uncertainties are
given at the 1$\sigma$ level, unless otherwise stated, and the photon indices are
parametrized as $dN/dE \propto E^{-\Gamma}$ with $\Gamma$ = $\alpha$+1
($\alpha$ is the spectral index).

\section{{\em Fermi}-LAT Data: Selection and Analysis}
\label{FermiData}

The {\em Fermi}-LAT  is a pair-conversion telescope operating from 20 MeV to
$>$ 300 GeV. It has a large peak effective area ($\sim$8000 cm$^{2}$ for 1
GeV photons), an energy resolution of typically $\sim$10\%, and a field of
view of about 2.4 sr with an angular resolution (68\% containment angle) better than 1$^{\circ}$ for energies above 1 GeV. Further
details about the {\em Fermi}-LAT are given in \citet{atwood09}. 

\begin{figure}
\centering
\includegraphics[width=7.5cm]{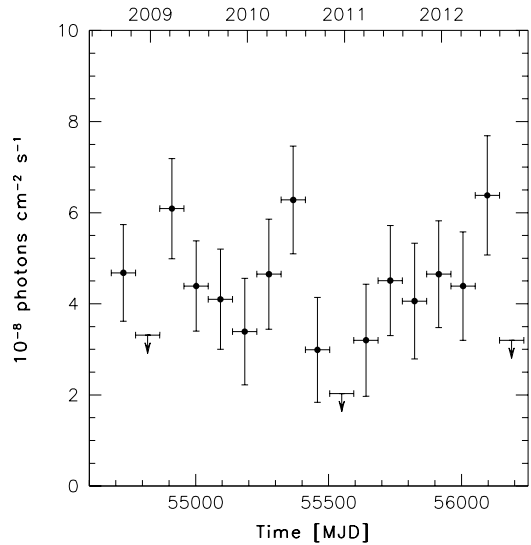}
\caption{Integrated flux light curve of PKS 1502$+$036 obtained in the 0.1--100 GeV energy range during 2008 August 4--2012 November 4 with 3-month time bins. Arrows refer to 2$\sigma$ upper limits on the source flux. Upper limits are computed when TS $<$ 10.}
\label{Fig1}
\end{figure}

The LAT data reported in this paper were collected from 2008 August 4  (MJD
54682) to 2012 November 4 (MJD 56235). During this time, the LAT instrument
operated almost entirely in survey mode. The analysis was performed with the \texttt{ScienceTools} software package version v9r27p1. The LAT data
were extracted within a $10^{\circ}$ region of interest centred at the location of PKS 1502$+$036. Only events belonging to the `Source' class were used. In addition, a cut on the
zenith angle ($< 100^{\circ}$) was also applied to reduce contamination from
the Earth limb $\gamma$-rays, which are produced by cosmic rays interacting with the upper atmosphere. 
The spectral analysis was performed with the instrument response functions \texttt{P7SOURCE\_V6} using an unbinned maximum-likelihood method implemented
in the Science tool \texttt{gtlike}. A Galactic diffuse emission model and isotropic component, which is the sum of
an extragalactic and residual cosmic ray background were used to model the background\footnote{http://fermi.gsfc.nasa.gov/ssc/data/access/lat/Background\\Models.html}. The normalizations of both components in the background model were allowed to vary freely during the spectral fitting. 

We evaluated the significance of the $\gamma$-ray signal from the sources by
means of the maximum-likelihood test statistic TS = 2$\Delta$log(likelihood) between models with
and without a point source at the position of PKS 1502$+$036 \citep{mattox96}. The source model used in
\texttt{gtlike} includes all the point sources from the 2FGL catalogue that
fall within $15^{\circ}$ from PKS 1502$+$036. The spectra of these sources
were parametrized by power-law (PL) functions, $dN/dE \propto$ $(E/E_{0})^{-\Gamma}$,
except for 2FGL J1504.3$+$1029 and 2FGL J1512.8$-$0906, for which we used a
log-parabola (LP), $dN/dE \propto$ $E/E_{0}^{-\alpha-\beta \, \log(E/E_0)}$ \citep[]{landau86, massaro04}, as in the 2FGL catalogue. 
A first maximum-likelihood analysis was performed to remove from the model the sources having
TS $<$ 10 and/or the predicted number of counts based on the fitted model
$N_{pred} < 3 $. A second maximum-likelihood analysis was performed
on the updated source model. The fitting procedure has been performed with the sources within 10$^{\circ}$ from
PKS 1502$+$036 included with the
normalization factors and the photon indices left as free parameters. For the
sources located between 10$^{\circ}$ and 15$^{\circ}$ from our target, we kept the
normalization and the photon index fixed to the values of the 2FGL catalogue.

The $\gamma$-ray point source localization by means of the \texttt{gtfindsrc}
tool applied to the $\gamma$-rays extracted during the 51 months of
observation (MJD 54682--56235) results in RA = 226.27$^{\circ}$, Dec. = 3.45$^{\circ}$
(J2000), with a 95\% error circle radius of 0.07$^{\circ}$, at an angular
separation of 0.01$^{\circ}$ from the radio position of PKS 1502$+$036. This implies a strict spatial association with
the radio coordinates of the NLSy1 PKS 1502$+$036.
Integrating over the entire period considered, the fit yielded a TS = 305 in the 0.1--100 GeV energy range, with an integrated
average flux of (4.0 $\pm$ 0.4)$\times$10$^{-8}$ photons cm$^{-2}$ s$^{-1}$
and a photon index of $\Gamma$ = 2.60 $\pm$ 0.06. The corresponding apparent
isotropic $\gamma$-ray luminosity is 7.8$\times$10$^{45}$ erg s$^{-1}$.
In order to test for curvature in
the $\gamma$-ray spectrum of PKS 1502$+$036, an alternative spectral model to the PL, a LP was used for the fit.
We obtain a spectral slope $\alpha$ = 2.45 $\pm$ 0.11 at the reference energy $E_0$ = 300 MeV, a curvature parameter around the peak $\beta$ = 0.10
$\pm$ 0.06, with a TS = 307 and an integrated average flux of (3.7 $\pm$ 0.4)$\times$10$^{-8}$ photons cm$^{-2}$ s$^{-1}$. We used a likelihood ratio test
to check the PL model (null hypothesis) against the LP model (alternative hypothesis). These
values may be compared
by defining the curvature test statistic TS$_{\rm curve}$=TS$_{\rm LP}$--TS$_{\rm PL}$=2 corresponding to
$\sim$1.4$\sigma$, meaning that we have no statistical evidence of a curved spectral shape.

Fig.~\ref{Fig1} shows the $\gamma$-ray light curve for the period 2008
August 4--2012 November 4 using a PL model and 3-month time bins. For each time bin, the spectral parameters
of PKS 1502$+$036 and all sources within 10$^{\circ}$ from it were frozen to the value resulting
from the likelihood analysis over the entire period. If TS $<$ 10, the value of
the fluxes were replaced by the 2$\sigma$ upper limits, calculated using the
profile method \citep[see e.g.][]{nolan12}. The systematic
uncertainty in the flux is energy dependent: it amounts to $10\%$ at 100 MeV, decreasing to
$5\%$ at 560 MeV, and increasing again to $10\%$ above 10 GeV
\citep{ackermann12}. Despite integrating over three months, the source was not
always detected. We test for flux variability calculating the variability index as
defined in \citet{nolan12}. We obtain a TS$_{\rm var}$ = 25.2 for 16
degrees of freedom, below the value of 32 defined to identify variable sources at
a 99\% confidence level. Thus, no significant variability was
detected in $\gamma$-rays, as reported also in the 2FGL catalogue over the
first 24 months of {\em Fermi} operations for this source
\citep{nolan12}. On weekly time-scales, the source was detected with TS $>$ 10
only a few times, with a peak flux of (18 $\pm$ 6)$\times$10$^{-8}$ photons
cm$^{-2}$ s$^{-1}$ in the 0.1--100 GeV energy range during 2012 July 7--13.

Leaving the photon index free to vary in the fits for three 3-month periods of
highest activity (flux $>$ 6$\times$10$^{-8}$ photons cm$^{-2}$
s$^{-1}$ in the 0.1--100 GeV energy range) and $E_{0}$ fixed to 548 MeV, as in the 2FGL catalogue, the fits result in photon index values $\Gamma$ = 2.57 $\pm$ 0.16, $\Gamma$ = 2.44 $\pm$ 0.15 and $\Gamma$ = 2.54
$\pm$ 0.18, in chronological order. No conclusive evidence for spectral evolution has been detected.
By means of the \texttt{gtsrcprob} tool, we estimated that
the highest energy photon emitted by PKS 1502$+$036 (with probability $>$ 80\% of being associated with the source) was
observed on 2009 October 30 at a distance of 0.05$^{\circ}$ from PKS 1502$+$036 with an energy of 19.6 GeV.

\begin{table*}
\caption{Log and fitting results of {\em Swift}/XRT observations of PKS 1502$+$036 using a power-law model with $N_{\rm H}$ fixed to Galactic
  absorption.}
\begin{center}
\begin{tabular}{cccccc}
\hline \hline
\multicolumn{1}{c}{Date (UT)} &
\multicolumn{1}{c}{MJD} &
\multicolumn{1}{c}{Net exposure time} &
\multicolumn{1}{c}{Net count rate} &
\multicolumn{1}{c}{Photon index} &
\multicolumn{1}{c}{Flux 0.3--10 keV$^{\rm a}$} \\
\multicolumn{1}{c}{} &
\multicolumn{1}{c}{} &
\multicolumn{1}{c}{(sec)} &
\multicolumn{1}{c}{($\times$10$^{-3}$ cps)} &
\multicolumn{1}{c}{($\Gamma$)} &
\multicolumn{1}{c}{($\times$10$^{-13}$ erg cm$^{-2}$ s$^{-1}$)} \\
\hline
2009-July-25 & 55037 & 4681 & 7.5 $\pm$ 1.3 & $1.2 \pm 0.4$ & $4.9 \pm 1.1$ \\
2012-Apr-25  & 56042 & 4807 & 7.7 $\pm$ 1.3 & $1.7 \pm 0.4$ & $4.0 \pm 0.9$ \\
2012-May-25  & 56072 & 4635 & 8.0 $\pm$ 1.3 & $1.9 \pm 0.4$ & $3.7 \pm 0.9$ \\
2012-June-25 & 56103 & 5142 & 8.4 $\pm$ 1.3 & $2.2 \pm 0.4$ & $3.5 \pm 0.7$ \\
2012-Aug-07/08 & 56146/7 & 4925 & 10.8 $\pm$ 1.5 & $2.2 \pm 0.3$ & $4.0 \pm 0.6$ \\
\hline
\hline
\end{tabular}
\end{center}
$^{\rm a}$Observed flux
\label{XRT}
\end{table*}

\section{{\em Swift} Data: Analysis and Results}
\label{SwiftData}

The {\em Swift} satellite \citep{gehrels04} performed six observations
of PKS 1502$+$036 between 2009 July and 2012 August. The observations were
performed with all three onboard instruments: the X-ray Telescope \citep[XRT;][0.2--10.0 keV]{burrows05}, the Ultraviolet/Optical Telescope \citep[UVOT;][170--600 nm]{roming05} and the Burst Alert Telescope \citep[BAT;][15--150 keV]{barthelmy05}.

The hard X-ray flux of this source turned out to be below the sensitivity of the BAT
instrument for such short exposures and therefore the data from this instrument will not be used.
Moreover, the source was not present in the {\em Swift} BAT 70-month hard X-ray catalogue \citep{baumgartner13}.

The XRT data were processed with standard procedures (\texttt{xrtpipeline
  v0.12.6}), filtering, and screening criteria by using the \texttt{HEAsoft}
package (v6.12). The data were collected in photon counting mode in all the observations. The source count rate
was low ($<$ 0.5 counts s$^{-1}$); thus pile-up correction was not
required. The data collected during 2012 August 7 and 8 were
summed in order to have enough statistics to obtain a
good spectral fit. Source events were extracted from a circular region with a radius of
20 pixels (1 pixel $\sim$ 2.36$"$), while background events were extracted
from a circular region with radius of 50 pixels far away from the source
region. Ancillary response files were generated with \texttt{xrtmkarf}, and
account for different extraction regions, vignetting and point spread function
corrections. We used the spectral redistribution matrices v013 in the
Calibration data base maintained by \texttt{HEASARC}. Considering the low number of
photons collected ($<$ 200 counts) the spectra were rebinned with a
minimum of 1 count per bin and we used Cash statistics \citep{cash79}. We have fitted the spectrum with an absorbed power-law using the photoelectric absorption model
\texttt{tbabs} \citep{wilms00}, with a neutral hydrogen column density fixed
to its Galactic value \citep[3.93$\times$10$^{20}$
cm$^{-2}$;][]{kalberla05}.
The results are reported in
Table~\ref{XRT}. In the past, the source has not been detected in X-rays
during the {\em ROSAT} all-sky survey, with an upper limit of $<$ 2.1$\times$10$^{-12}$ erg cm$^{-2}$
s$^{-1}$ in the 0.1--2.4 keV energy range \citep{yuan08}. PKS 1502$+$036 seems
to have been observed by {\em Swift}/XRT in a relatively bright state on 2009 July 25
in the 0.3--10 keV energy range. In addition a
possible moderate change of the photon index was observed during
2009--2012 together with an increase of the count rate leading up to 2012 August 7--8, but no
conclusive evidence can be drawn on consideration of the large uncertainties. 

During the {\em Swift} pointings, the UVOT instrument observed PKS 1502$+$036
in all its optical ($v$, $b$ and $u$) and UV ($w1$, $m2$ and $w2$) photometric
bands \citep{poole08,breeveld10}. We analysed the data using the
\texttt{uvotsource} task included in the \texttt{HEAsoft} package. Source
counts were extracted from a circular region of 5 arcsec radius centred on the source, while background
counts were derived from a circular region of 10 arcsec radius in the  source
neighbourhood. We combined the images collected on 2012 August 7 and 8 in
  order to achieve better statistics. The observed magnitudes are reported in
Table~\ref{UVOT}. Upper limits are calculated using the UVOT photometric
system when the analysis provided a significance of detection $<$ 3$\sigma$.

Magnitudes in optical filters are comparable within the uncertainties
between the different {\em Swift} observations, and compatible with the values
observed in the past \citep[$V$=18.62--18.82; $B$=19.11--19.27; $U$=18.58--18.68;][]{wills78}. Recently,
optical intraday variability with amplitude as large as $\sim$10\% has been
reported for PKS 1502$+$036 by \citet{paliya13}. No evident infrared intraday
variability has been found with {\em Wide-field Infrared Survey Explorer (WISE)} data, but a variation of 0.1--0.2 mag corrected for measurement errors during two epochs separated by about 180 d has been reported in \citet{jiang12}.

\begin{table*}
\caption{Results of the {\em Swift}/UVOT data for PKS 1502$+$036. Upper limits
are calculated when the analysis provided a significance of detection $<$ 3$\sigma$.}
\begin{center}
\begin{tabular}{cccccccc}
\hline \hline
\multicolumn{1}{c}{Date (UT)} &
\multicolumn{1}{c}{MJD} &
\multicolumn{1}{c}{$v$} &
\multicolumn{1}{c}{$b$} &
\multicolumn{1}{c}{$u$} &
\multicolumn{1}{c}{$w1$} &
\multicolumn{1}{c}{$m2$} &
\multicolumn{1}{c}{$w2$} \\
\hline
2009-July-25 & 55037 &  $>$ 18.78 &  19.31$\pm$0.28 & 18.68$\pm$0.23 & 18.57$\pm$0.18 & 18.25$\pm$0.16 & 18.49$\pm$0.12 \\
2012-Apr-25  & 56042 &  18.65$\pm$0.33 &  $>$ 19.80 & 18.98$\pm$0.27 & 18.98$\pm$0.22 & 18.85$\pm$0.10 & 18.58$\pm$0.11 \\
2012-May-25  & 56072 &  18.72$\pm$0.31 &  19.28$\pm$0.22 & 18.74$\pm$0.06 & 18.32$\pm$0.14 & 18.60$\pm$0.17 & 18.35$\pm$0.10 \\
2012-June-25 & 56103 &  $>$ 18.76 &  19.12$\pm$0.27 & 18.67$\pm$0.27 & 18.51$\pm$0.08 & 18.25$\pm$0.15 & 18.62$\pm$0.13 \\
2012-Aug-07/08 & 56146/7 & 19.39$\pm$0.30 & 20.12$\pm$0.28 & 18.90$\pm$0.14 & 18.79$\pm$0.10 & 18.61$\pm$0.08 & 18.65$\pm$0.06 \\
\hline
\hline
\end{tabular}
\end{center}
\label{UVOT}
\end{table*}

\section{Catalina Real-Time Transient Survey}

The source has been monitored by the CRTS \citep[http://crts.caltech.edu;][]{drake09, djorgovski11},
using the 0.68 m Schmidt telescope at Catalina Station, AZ, and an unfiltered
CCD.  The typical cadence is to obtain four exposures separated by 10 min in a
given night; this may be repeated up to four times per lunation, over a period of $\sim$6--7 months each year, while the field is
observable.  Photometry is obtained using the standard Source-Extractor
package \citep{bertin96}, and roughly calibrated to the $V$ band in terms of the
magnitude zero-point. During the CRTS monitoring, the source showed a variability amplitude of 1 mag, changing between 18.7 and 17.7 mag.

\section{Radio Data: Analysis and Results}\label{RadioData}

\subsection{VLA and VLBA data}

Multi-epoch and multifrequency observations of PKS\,1502$+$036 were
carried out with both the VLA and VLBA between 1999 and 2007 as part of the
monitoring campaign of the `bright' HFP sample
\citep{dallacasa00}. Results of the observations performed between
1999 and 2003 were presented in
\citet{dallacasa00,tinti05,orienti06,orienti07}. To study the proper
motion, we complemented our VLBA observations with the 
15 GHz VLBA data from the \texttt{MOJAVE} programme\footnote{The \texttt{MOJAVE} data archive is 
maintained at http://www.physics.purdue.edu/MOJAVE.} \citep{lister09}
and additional 15 GHz archival VLBA observations. Logs of VLA and VLBA
observations are reported in Table~\ref{vla-logs}.

Simultaneous VLA observations were performed in $L$ band 
(with the two intermediate frequencies, IFs, centred at 1.415 and 1.665 GHz), 
$C$ band (with the two IFs centred at 4.565 and 4.935 GHz), 
$X$ band (with the two IFs centred at 8.085 and 8.465 GHz), 
$U$ band (14.940 GHz), and $K$ band (22.460 GHz). At each frequency the target
source was observed for about 1 min, cycling through 
frequencies. During each run, the primary calibrator 3C\,286 was
observed for about 3 min at each frequency. The data reduction was 
carried out following the standard procedures for the VLA implemented
in the National Radio Astronomy Observatory (NRAO) Astronomical Image
Processing System (\texttt{AIPS}) package. The flux density at each frequency 
was measured on the final image produced after a few phase-only 
self-calibration iterations. In the $L$ band, it was generally necessary 
to image a few confusing sources falling within the primary beam. 
The target source is unresolved at any frequency even with the VLA
A-configuration. During one epoch, strong radio frequency interference at 1.665 GHz was present, and a measurement of the flux density was not
possible. Uncertainties on the determination of the absolute flux density
scale are dominated by amplitude errors, which are about 3\% in $L$, $C$, and $X$ bands, about 5\% in $U$ band, and about 10\% in $K$ band.

Simultaneous VLBA observations were performed in 2002 January at 15.3
and 22 GHz, and in 2006 July at 1.7 GHz, 2.3 GHz
($S$ band), 5.0 GHz, 8.4 GHz and 15.3
GHz. The correlation was performed at the VLBA hardware correlator 
in Socorro and the data reduction was carried out with the NRAO
\texttt{AIPS} package. After the application of system temperatures and
antenna gains, the amplitudes were checked using data on 4C 39.25 (J0927+3902), which was used also as bandpass calibrator. The error in the absolute flux density scale is generally within
3\%-10\%, being highest in value at the highest frequency. \\
The final images were obtained after a number of self-calibration
iterations. Amplitude self-calibration was applied only once at the
end of the process, using particular care; the solution interval (30
min) was chosen to be longer than the scan-length to remove residual
systematic errors and fine tune the flux density scale, but not to
force the individual data points to follow the model. \\
Additional archival data have been considered to improve the time
sampling. A priori amplitude calibration was derived using measurements of
the system temperature and the antenna gains. Uncertainties of the amplitude
calibrations were assumed to be 5\%. For \texttt{MOJAVE} data, we imported the fully calibrated {\it uv} data sets
\citep{lister09}.

\begin{table*}
\caption{Logs of VLA and multifrequency VLBA observations. Column 1: observation
  date in UT; Column 2: observation date in MJD; Column 3: observation code; Column 4: array; Column 5: observing bands;
  Column 6: code used in this paper; Column 7: reference:
  1=\citet{dallacasa00}; 2=\citet{tinti05}; 3=\citet{orienti07};
  4=this paper; 5=\citet{orienti06}.}
\begin{center}
\begin{tabular}{ccccccc}
\hline
\hline
Date (UT) & MJD & Observation code&Array&Bands&Code&Reference\\
\hline
1999-Sep-25 & 51446 & AD428 & VLA-A & $L$,$C$,$X$,$U$,$K$&{\it a}& 1 \\
2002-July-03 & 52458 & AT275 & VLA-B & $L$,$C$,$X$,$U$,$K$&{\it b}& 2 \\
2003-Sep-13 & 52895 & AD488 & VLA-AnB &$L$,$C$,$X$,$U$,$K$&{\it c}& 3 \\
2007-Apr-14 & 54204 & AO210 & VLA-D   &$L$,$C$,$X$,$K$&{\it d}& 4 \\
\hline
2001-Oct-06 & 52188 & BF068 & VLBA\tablenotemark{a}  & $U$  &{\it e}& 4 \\
2002-Jan-11 & 52285 & BD077 & VLBA    & $X$,$U$&{\it f}& 5 \\
2002-Sep-26 & 52481 & RDV35 & VLBA    & $X$ &{\it g}& 4 \\ 
2005-Dec-07 & 53711 & BF075 & VLBA\tablenotemark{a}  & $U$  &{\it h}& 4 \\
2006-June-26 & 53912 & BF075 & VLBA\tablenotemark{a}& $U$  &{\it i}& 4 \\
2006-July-22 & 53938 & BO025 & VLBA    & $L$,$S$,$C$,$X$,$U$&{\it l}& 4 \\
2010-Oct-15 & 55484 & BL149 & VLBA\tablenotemark{b}& $U$    &{\it m}& 4 \\
2010-Oct-25 & 55494 & BL149 & VLBA\tablenotemark{b}& $U$    &{\it n}& 4 \\
2011-May-06 & 55687 & BL149 & VLBA\tablenotemark{b}& $U$    &{\it o}& 4 \\
2012-Jan-02 & 55928 & BL178 & VLBA\tablenotemark{b}& $U$    &{\it p}& 4 \\
2012-Sep-27 & 56197 & BL178 & VLBA\tablenotemark{b}& $U$    &{\it q}& 4 \\     
\hline
\hline
\end{tabular}
\end{center}
$^{\rm a}$archival VLBA data; $^{\rm b}$data from the \texttt{MOJAVE} programme
\label{vla-logs}
\end{table*}

\begin{table*}
\caption{VLA flux density. Column 1: code from Table \ref{vla-logs};
  Columns 2-9: flux density at 1.4, 1.7, 4.5, 4.9, 8.1, 8.4, 15.3, 22.2
  GHz, respectively; Column 10: peak frequency; Column 11: peak flux density; Columns 12 and 13: spectral index below and above the
  peak frequency, respectively; Column 14: variability index (see
  Section \ref{spectra}).}  
\begin{center}
\begin{tabular}{cccccccccccccc}
\hline
\hline
Code&$S_{\rm 1.4}$&$S_{\rm 1.7}$&$S_{\rm 4.5}$&$S_{\rm 4.9}$&$S_{\rm
  8.1}$&$S_{\rm 8.4}$&$S_{\rm 15.3}$&$S_{\rm 22.2}$&$\nu_{\rm p}$&$S_{\rm p}$&$\alpha_{b}$&$\alpha_{a}$&$V$\\
 &(mJy)&(mJy)&(mJy)&(mJy)&(mJy)&(mJy)&(mJy)&(mJy)&(GHz)&(mJy)& & &\\
\hline
{\it a} & 455 & 547 & 921 & 929 & 901 & 895 & 830 & 738 & 7.7$\pm$0.8& 940 &$-$1.7&0.2&107\\
{\it b} & 413 & 465 & 740 & 741 & 721 & 710 & 665 & 567 & 7.2$\pm$0.7& 745 &$-$0.5&0.2&3\\
{\it c} & 382 & 428 & 608 & 620 & 620 & 610 & 515 & 468 & 6.5$\pm$0.6& 622 &$-$0.4&0.3&122\\
{\it d} & 332 & --  & 679 & 692 & 720 & 720 & -- & 611 & 8.8$\pm$0.9& 737 &$-$0.5&0.2&38\\
\hline
\hline
\end{tabular}
\end{center}
\label{vla-flux}
\end{table*}

\begin{table*}
\caption{Multifrequency VLBA flux density. Column 1: code from Table \ref{vla-logs};
  Columns 2-6: flux density at 1.7, 2.3, 5.0, 8.4, 15.3 GHz,
  respectively; Columns 7 and 8: peak frequency and peak flux density,
  respectively (see Section \ref{spectra});
  Columns 9 and 10: spectral index below and above the peak frequency,
  respectively.} 
\begin{center}
\begin{tabular}{cccccccccc}
\hline
\hline
Code&$S_{\rm 1.7}$&$S_{\rm 2.3}$&$S_{\rm 5.0}$&$S_{\rm 8.4}$&$S_{\rm
  15.3}$&$\nu_{\rm p}$&$S_{\rm p}$&$\alpha_{\rm b}$&$\alpha_{\rm a}$\\
 &(mJy)&(mJy)&(mJy)&(mJy)&(mJy)&(GHz)&(mJy)& & \\
\hline
{\it f} & --  & -- & -- & 630  & 566 & -- & -- & -- & -- \\
{\it g} & --  & -- & -- & 800  & --  & -- & -- & -- & -- \\
{\it l} & 365 & 480 & 670 & 701 & 547 & 6.6$\pm$0.7& 704 &$-$0.6&0.4 \\
\hline
\hline
\end{tabular}
\end{center}
\label{vlba-flux}
\end{table*}

\subsection{OVRO}\label{ovro}

\begin{figure}
\centering
\includegraphics[width=7.5cm]{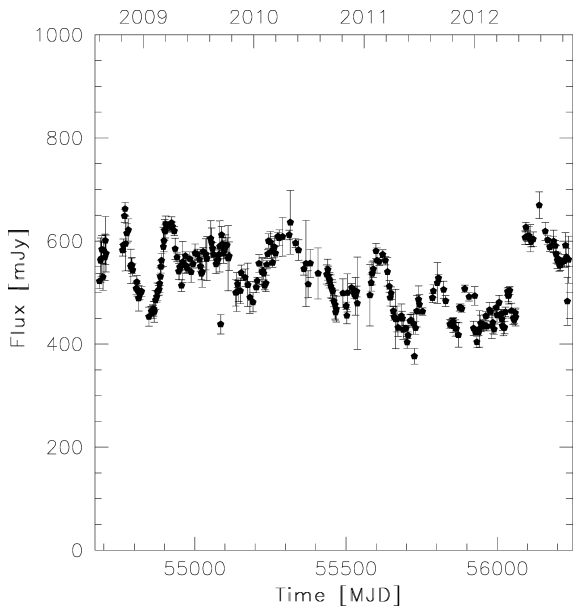}
\caption{15 GHz radio light curve of PKS\,1502$+$036 for the period 2008 August 7--2012 November 4 from the OVRO telescope.}
\label{Fig2}
\end{figure}

As part of an ongoing blazar monitoring programme, the OVRO 40 m radio telescope has observed PKS 1502$+$036
at 15~GHz regularly since the end of 2007 \citep{richards11}. This
monitoring programme includes over 1500 known and likely $\gamma$-ray loud
blazars above declination $-20^{\circ}$. The sources in this
programme are observed in total intensity twice per week with a 4~mJy
(minimum) and 3\% (typical) uncertainty in their flux densities. Observations are performed
with a dual-beam (each 2.5~arcmin FWHM) Dicke-switched system using
cold sky in the off-source beam as the reference. Additionally, the
source is switched between beams to reduce atmospheric variations. The
absolute flux density scale is calibrated using observations of
3C~286, adopting the flux density (3.44~Jy) from \citet{baars77}. 
This results in about a 5\% absolute scale uncertainty, which is not reflected
in the plotted errors. As shown in Fig.~\ref{Fig2}, PKS 1502$+$036 was highly variable at 15 GHz during the
OVRO monitoring, with a flux density spanning 376 mJy (at MJD 55726) to 669 mJy (at MJD 56139).

\subsection{Radio data: results}

\begin{figure}
\centering
\includegraphics[width=7.5cm]{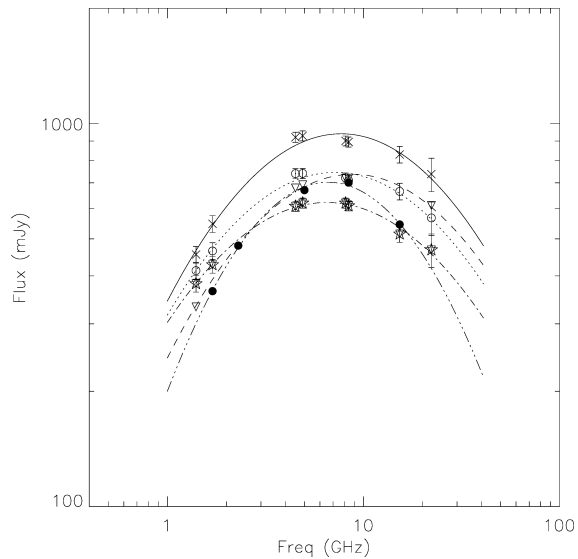}
\caption{Multi-epoch radio spectra of PKS\,1502$+$036 observed with VLA and
  VLBA during the monitoring campaign of the `bright' HFP sample.
 Epoch $a$: X symbol and solid line; epoch $b$:
  empty circle and dot line; epoch $c$: empty star and dot--dashed line;
  epoch $l$ (VLBA): filled circle and dash-and-three-dot line; epoch $d$: triangle
and dashed line. }
\label{variability}
\end{figure}

\subsubsection{The radio spectrum}
\label{spectra}

Simultaneous multifrequency VLA observations carried out at various epochs
showed substantial spectral and flux density variability. 
To determine the change in the spectral peak, we follow
the approach by \citet{orienti10} and we fit the simultaneous radio
spectrum at each epoch with a purely analytical function:

\begin{equation}
{\rm log \,S = a + log \,\nu \times (b + c\, log \,\nu) }
\end{equation}

\noindent where S is the flux density, $\nu$ the frequency, and a, b and c are
numeric parameters without any direct physical meaning. The best fits
to the spectra are shown in Fig.~\ref{variability}, 
and the derived peak frequencies at the various epochs are reported in
Tables~\ref{vla-flux} and \ref{vlba-flux}. Statistical errors derived
from the fit are not representative of the real uncertainty on the
estimate of the peak frequency. For this reason, we prefer to assume a
conservative uncertainty on the peak frequency of 10\%. In addition, we
computed the spectral index in the optically thick ($\alpha_{\rm b}$)
and thin part ($\alpha_{\rm a}$) of the spectrum. Both the spectral
shape and the spectral peak change randomly among the various observing epochs, 
similar to what is found in blazars \citep[e.g.][]{torniainen07} as
well as in radio-loud NLSy1s \citep[e.g.][]{yuan08,dammando12}. 

Following the approach of \citet{orienti07}, we estimated the spectral
variability by means of the variability index:

\begin{equation}
V = \frac{1}{m} \sum^{m}_{i=1} \frac{(S_{i} - \bar{S_{i}})^{2}}{\sigma_{i}^{2}},
\label{index}
\end{equation}

where $S_{i}$ is the flux density at the $i$th frequency
measured at one epoch, $\overline S_{i}$ is the mean value of the flux
density computed by averaging the flux density at the $i$th frequency
measured at all the available epochs, $\sigma_{i}$ is the root mean square (rms) 
on $S_{i} - \overline S_{i}$ and $m$ is the number of sampled frequencies. 
The variability index computed at the different epochs ranges between
3 and 122, indicating that this source is highly variable (Table~\ref{vla-flux}). We prefer to compute the variability index for each epoch 
instead of considering all the epochs together in order to potentially detect small outbursts. 

\subsubsection{Radio structure}

\begin{figure}
\centering
\includegraphics[width=7.0cm]{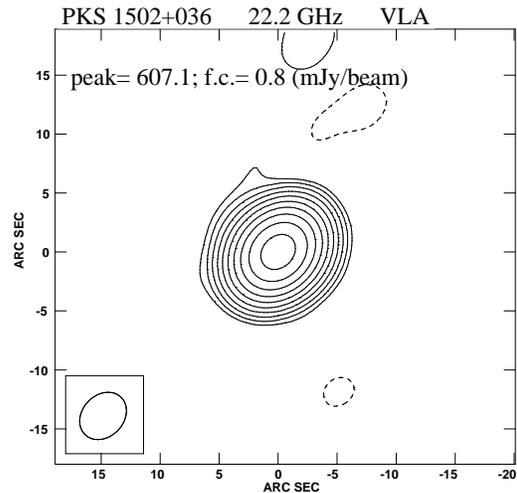}
\caption{VLA image at 22.2 GHz of PKS 1502$+$036. On the image, we provide the
  restoring beam, plotted in the bottom-left corner, the peak flux density in
  mJy/beam and the first contour (f.c.) intensity in mJy/beam, which is three
  times the off-source noise level. Contour levels increase by a factor of 2.}
\label{VLA}
\end{figure}

\begin{figure}
\centering
\includegraphics[width=7.0cm]{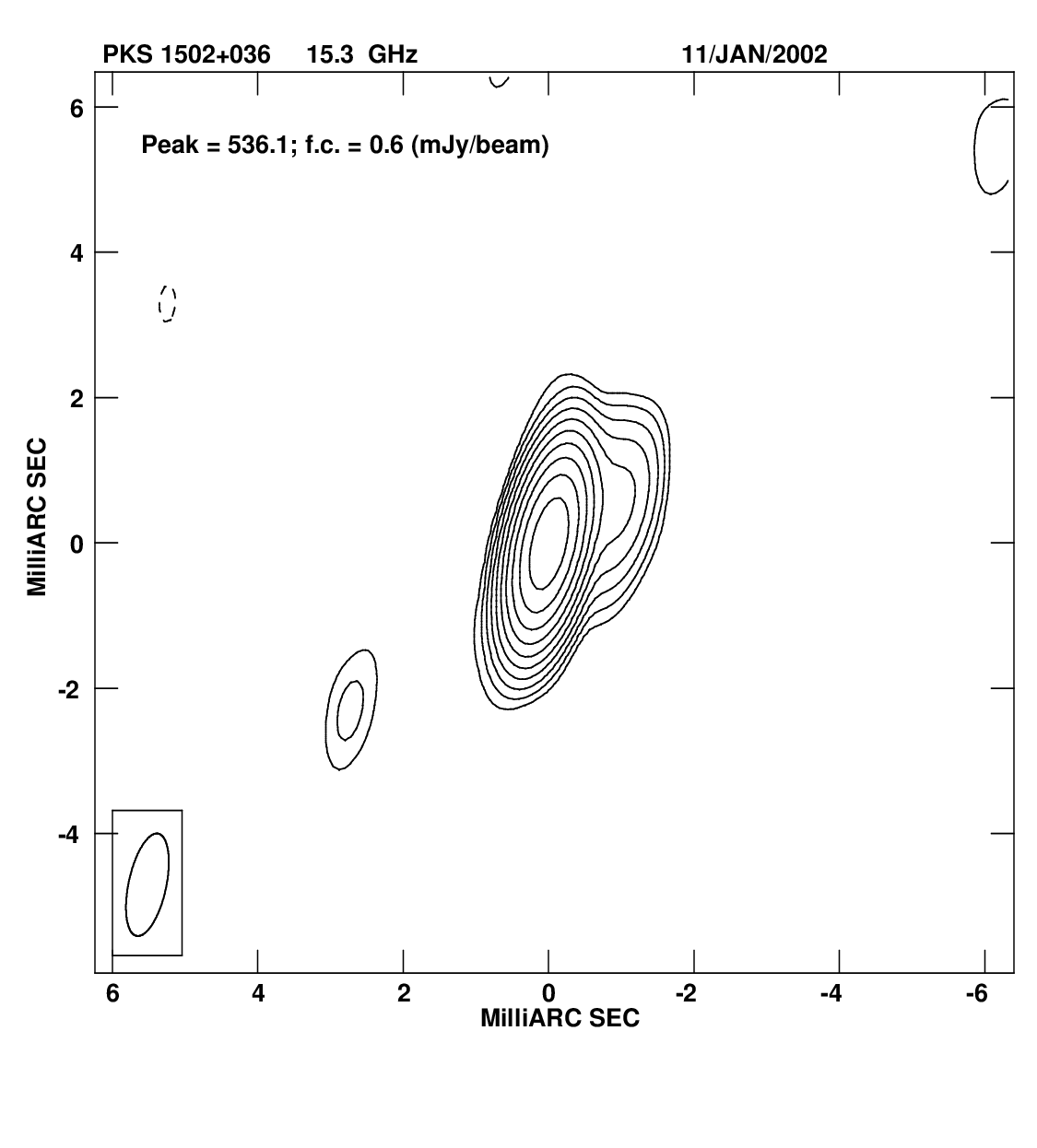}
\caption{VLBA image at 15.3 GHz of PKS 1502$+$036 collected on 2002 January 11. On the image, we provide the
  restoring beam, plotted in the bottom-left corner, the peak flux density in
  mJy/beam and the first contour (f.c.) intensity in mJy/beam, which is three
  times the off-source noise level. Contour levels increase by a factor of 2.}
\label{VLBA_2002}
\end{figure}

\begin{figure}
\centering
\includegraphics[width=7.0cm]{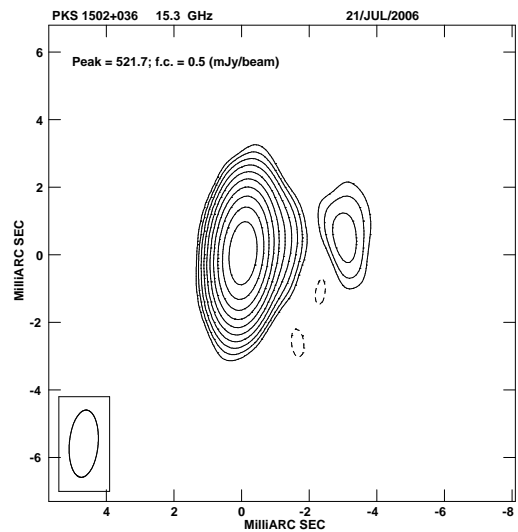}
\caption{VLBA image at 15.3 GHz of PKS 1502$+$036 collected on 2006 July 21. On the image, we provide the
  restoring beam, plotted in the bottom-left corner, the peak flux density in
  mJy/beam and the first contour (f.c.) intensity in mJy/beam, which is three
  times the off-source noise level. Contour levels increase by a factor of 2.}
\label{VLBA_2006}
\end{figure}

\begin{figure}
\centering
\includegraphics[width=7.0cm]{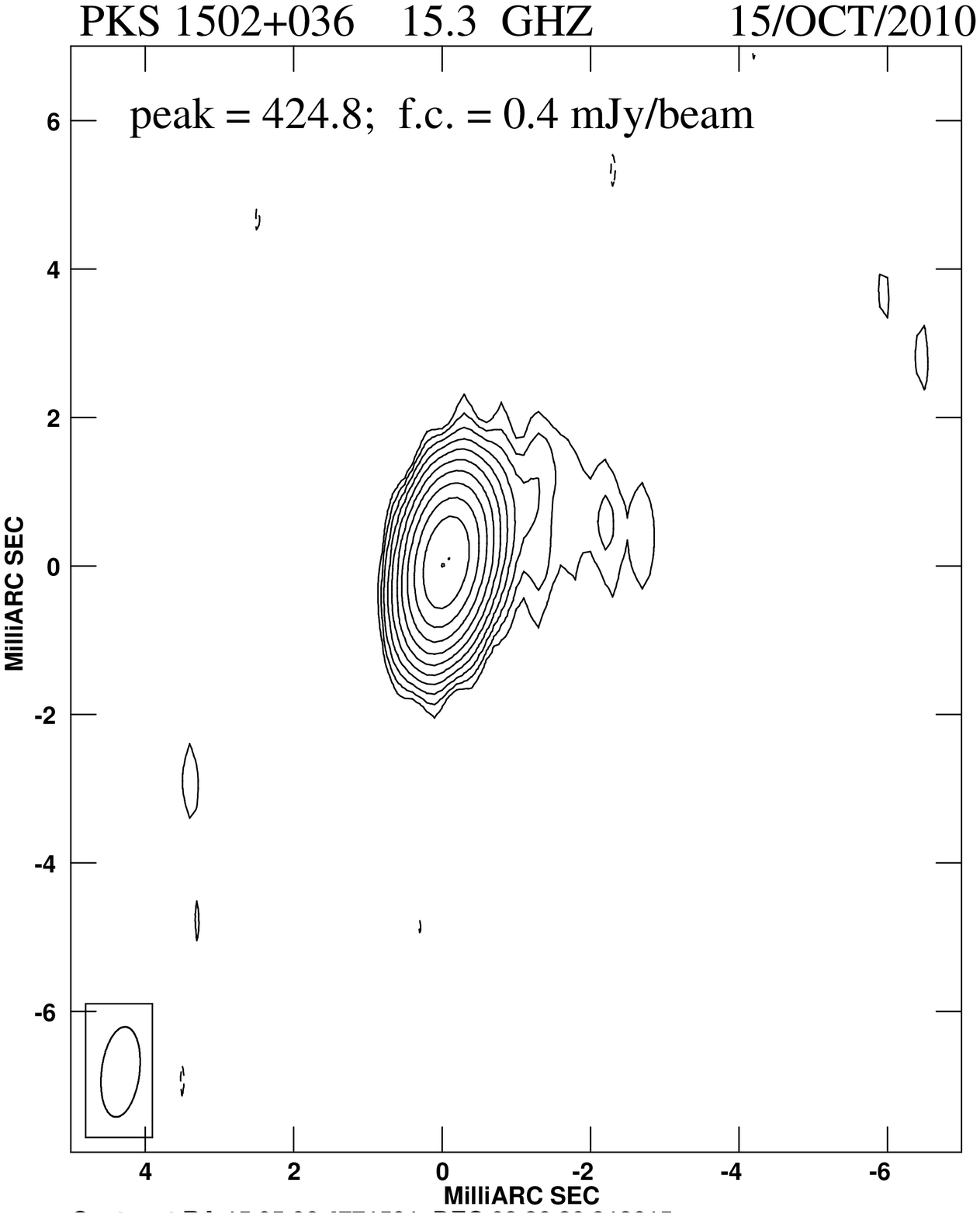}
\caption{VLBA image at 15.3 GHz of PKS 1502$+$036 collected on 2010 October 15. On the image, we provide the
  restoring beam, plotted in the bottom-left corner, the peak flux density in
  mJy/beam and the first contour (f.c.) intensity in mJy/beam, which is three
  times the off-source noise level. Contour levels increase by a factor of 2.}
\label{VLBA_2010}
\end{figure}

PKS\,1502$+$036 is unresolved on arcsecond scales typical of the VLA
(Fig.~\ref{VLA}). However, when imaged with the parsec scale resolution provided by VLBA observations, its radio structure is
marginally resolved and a second component seems to emerge from the core. In
the VLBA images at 15 GHz, the radio structure is clearly extended, suggesting a core-jet-like morphology (Figs~\ref{VLBA_2002}, \ref{VLBA_2006} and \ref{VLBA_2010}). The radio
emission is dominated by the core component, which is unresolved in the 15 GHz image with an angular size $<$ 0.4 mas. The jet-like
feature accounts for about 4\% of the total flux density in the observation
presented here, and it extends to about 3 mas (Fig.~\ref{VLBA_2010}). Since the
VLBA flux density is in good agreement with the nearly contemporaneous
measurements by the OVRO 40 m single dish, we can exclude significant
emission from extended, low surface brightness features.

\begin{figure}
\begin{center}
\includegraphics{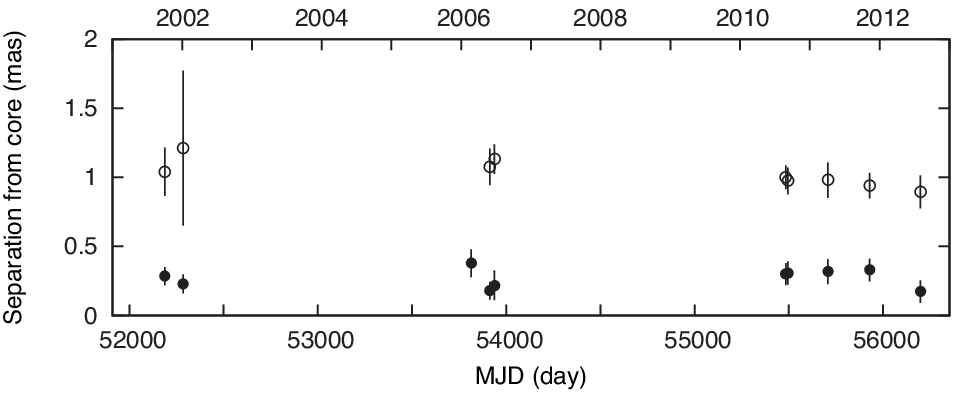}
\vspace{5cm}
\caption{Angular separation between the jet components (C1 and C2, full and empty circles respectively) and the core in
PKS\,1502$+$036.}
\label{proper}
\end{center}
\end{figure}

To investigate possible changes in the radio structure, we modelled
the performed visibility-based model fitting using the \texttt{DIFMAP}
software. 
The radio structure consists of a bright compact core plus weak 
extended emissions
westward and is well fitted with a point source (core) plus two (circular) Gaussian
components (`C1' and `C2'); we could not identify the third component for
the data of 2006 March 20 due to limited dynamic range. The results of image analyses are listed in Table~\ref{15ghz}.
The error in total flux density was determined from the uncertainties of the
amplitude calibration and image rms noise.  
The error in the measured separation between the core and the jet components 
was determined from the
root-sum-square of the uncertainties of the position measurements for C1 (or C2)
and the core. The uncertainty in the relative positions is estimated from

\begin{equation}
\Delta r = a/(S_{p}/\rm rms),
\label{error}
\end{equation}

\noindent where $a$ is the major axis of the deconvolved component, $S_{p}$ is its peak
flux density and rms is the 1$\sigma$ noise level measured on the
image plane \citep{polatidis03}. In case the errors estimated by
equation (\ref{error}) are unreliably small,  
we assume a more conservative value for $\Delta r$ that is 10$\%$
of the FWHM of the beam. \\
In contrast to what is found in the radio-loud NLSy1 SBS\,0846+513, where an
apparent superluminal motion has been detected \citep[$\beta \sim 10.9c$,][]{dammando12b}, no significant proper motion is detected for
the jet components of PKS\,1502$+$036 (Fig.~\ref{proper}).\\

\begin{table*}
\caption{Log and results of VLBA observations at 15 GHz.}
\label{15ghz}
\begin{center}
\begin{tabular}{lllllll}
\hline\hline
Date (UT) & MJD & Restoring beam & Image noise & VLBA total flux & C1 position & C2 position \\
 & &  (mas $\times$ mas) & (mJy/beam) & (mJy) & (mas) & (mas) \\
\hline
2001-Oct-06 & 52188 & $1.01 \times 0.41$ & 2.2 &
$439 \pm 22$ & $0.29 \pm 0.07$ & $1.04 \pm 0.17$ \\
2002-Jan-11 & 52285 & $1.06 \times 0.44$ & 5.2 & $565 \pm 29$ & 
$0.23\pm 0.07$ & $1.21 \pm 0.56$ \\
2006-Mar-20 & 53814 & $1.58 \times 0.62$ & 2.0 & $608 \pm 30$ & $0.38 \pm
0.10$ & -- \\
2006-June-27  & 53913 & $1.08 \times 0.36$ & 1.1 &
$553 \pm 28$ & $0.18 \pm 0.06$ & $1.08 \pm 0.13$ \\
2006-July-21 & 53937 & $1.60 \times 0.71$ & 0.8 & $547 \pm 27$ &
$0.22\pm 0.11$ & $1.13 \pm 0.11$ \\ 
2010-Oct-15 & 55484 & $1.22 \times 0.51$ & 0.2 &
$504 \pm 25$ & $0.30 \pm 0.08$ & $1.00 \pm 0.09$ \\
2010-Oct-25 & 55494 & $1.33 \times 0.52$ & 0.2 &
$487 \pm 24$ & $0.31 \pm 0.08$ & $0.97 \pm 0.10$ \\
2011-May-26 & 55707 & $1.32 \times 0.63$ & 0.3 &
$429 \pm 21$ & $0.32 \pm 0.09$ & $0.98 \pm 0.13$ \\
2012-Jan-02 & 55928 & $1.26 \times 0.53$ & 0.2 &
$444 \pm 22$ & $0.33 \pm 0.08$ & $0.94 \pm 0.09$ \\
2012-Sep-27 & 56197 & $1.23 \times 0.53$ & 0.3 & $588 \pm 29$ &
$0.17 \pm 0.08$ & $0.90 \pm 0.12$ \\
\hline\hline
\end{tabular}
\end{center}
\end{table*}
 
\section{Discussion and conclusions}

The detection of four radio-loud NLSy1s in $\gamma$-rays during the first year of {\em
  Fermi}-LAT operation \citep{abdo09} has stimulated the interest in this class
of AGN. Currently the search for new $\gamma$-ray emitting NLSy1s has led to only one
new source detected at high significance \citep{dammando12}. So far most of the
attention has been directed towards the two NLSy1s that showed $\gamma$-ray
flaring activity, PMN J0948$+$0022 and SBS 0846$+$513 \citep{foschini11, dammando12}. 

PKS 1502$+$036 showed no significant $\gamma$-ray variability during 2008 August 4--2012 November 4, with the 0.1--100 GeV
flux ranging between (3--7)$\times$10$^{-8}$ photons cm$^{-2}$ s$^{-1}$ using
3-month time bins. The average LAT spectrum accumulated over 51 months is well
described by a power-law with a photon index of $\Gamma$ = 2.60 $\pm$ 0.06,
which is similar to the mean value observed for flat spectrum radio quasars (FSRQs) during the first two years of {\em
  Fermi}-LAT operation \citep[$\Gamma$ = 2.42 $\pm$ 0.17;][]{ackermann12b}. 

\noindent In the same way, the average apparent isotropic $\gamma$-ray
luminosity of PKS 1502$+$036 is 7.8$\times$10$^{45}$ erg
s$^{-1}$ in the 0.1--100 GeV range, a typical value for a FSRQ
\citep[e.g.][]{grandi12}, and it is in agreement with what was found for SBS 0846$+$513 and PMN
J0948$+$0022 \citep[see discussion in][]{dammando12}. This suggests the
presence of a relativistic jet with Doppler factors ($\delta$) as large as in blazars. Modelling of the spectral
energy distribution (SED) of the radio-loud NLSy1s detected by {\it Fermi}-LAT
indicates Doppler factors larger than 10 \citep{abdo09,dammando12}, and in particular
$\delta \sim 18$ for PKS\,1502$+$036. However, the study of the
proper motion could not detect any significant motion for this
source, while the radio spectral variability and the one-sided structure
further require the presence of boosting effects in a relativistic
jet. 

The 15 GHz light curve (Sec.~\ref{ovro}) is highly variable. In particular, two outburst episodes are clearly present:
the first peaks at MJD 54923, and the second peaks at MJD 56139. These outbursts seem to take place when a flux enhancement
  is also observed in $\gamma$-rays. Unfortunately, the statistics in
  $\gamma$-rays are not adequate to allow a study of the variability on
  time-scales short enough to be related to the radio monitoring, while in the
  optical and UV bands the time sampling
was too poor. For this reason, we estimate the variability time-scale $\Delta
t$ on the basis of the radio data, similarly to \citet{valtoja99}. For $\Delta t$, we assume the time
  interval of the flux density variation to be between the minimum and maximum flux density of a single outburst $| \Delta S|$. This assumption implies that the minimum flux density
  corresponds to a stationary underlying component and the variation is due to
  a transient component. Taking into consideration the time dilation due to
  the cosmological redshift we find that the intrinsic time lag is $\Delta \tau = \Delta t/(1+z)$, while the intrinsic flux density
  variation at the observed frequency is $| \Delta S_{i}| =| \Delta S| \times (1+z)^{1-\alpha}$. The brightness temperature at a given frequency is

\begin{equation}  
 T_{\rm B}{\rm (\nu)} = \frac{1}{2k} \frac{S{\rm (\nu)}}{\Omega} \left( \frac{c}{\nu} \right)^{2},
\label{tb}
\end{equation}

\noindent where $k$ is the Boltzmann constant, $\Omega$ is the solid angle of the emitting region, and $c$ is the speed of light. Using
the causality principle, we can determine the angular size of the region
responsible for the outburst 

\begin{equation}
\theta = \frac{c \Delta t}{(1+z)} \frac{(1+z)^{2}}{\,D_{L}}\;\; .
\label{theta}
\end{equation}

\noindent Knowing that the solid angle is 

\begin{equation}
\Omega = \frac{\pi}{4} \theta^{2} \;, 
\label{omega}
\end{equation}

\noindent we derive the rest-frame brightness temperature from

\begin{equation}
T'_{B} = \frac{2}{\pi k} \frac{|\Delta S| D_{L}^{2}}{\Delta t^{2} \nu^{2}
  {\rm (} 1+z {\rm )^{1+ \alpha}}}  \;\; .
\label{tbrest}
\end{equation}

\noindent During the two outburst episodes we have flux density variability
$\Delta S =$ 185 and 216 mJy, respectively, while $\Delta t=$ 75 and
85 d, respectively. If in
equation~(\ref{tbrest}) we consider these values, and we assume $\alpha=0.3$, i.e. the average value obtained by fitting the
optically thin spectrum (Tables \ref{vla-flux} and \ref{vlba-flux}), 
we obtain T$'_{B}$ $\sim$ 2.5$\times$10$^{13}$ K, which exceeds the value derived for the Compton catastrophe. Assuming
that such a high value is due to Doppler boosting, we can estimate the
variability Doppler factor $\delta_{\rm var}$, by means of

\begin{equation}
\delta_{var} = \left( \frac{T'_{B}}{T_{int}} \right)^{1/(3+ \alpha)},
\label{dopplervar}
\end{equation}

\noindent where $T_{int}$ is the intrinsic brightness temperature. Assuming a
typical value $T_{int}$= 5$\times$10$^{10}$ K, as derived by
e.g.~\citet{readhead94,hovatta09,lahteenmaki99}, we obtain $\delta_{var} =$ 6.6.

Another way to derive a lower limit to the Doppler factor is by means of the
jet/counter-jet brightnesses ratio. Assuming that both the jet
  and counter-jet have the same intrinsic power, the different brightness observed can be related to Doppler boosting by means of

\begin{equation}
R = \frac{B_{\rm j}}{B_{\rm cj}} = \left( \frac{1 + \beta {\rm cos}\theta}{1 - \beta {\rm cos}\theta}\right)^{2 + \alpha},
\label{ratio}
\end{equation}

where $B_{\rm j}$ and $B_{\rm cj}$ are the jet and counter-jet
brightness, respectively, $\theta$ is the viewing angle, $\beta$ is the
bulk velocity in terms of the speed of light. We prefer to compare the
surface brightness instead of the flux density because the jet has a smooth
structure without clear knots. We measured the jet brightness as 20 mJy/beam in the 15 GHz image. In the case of the
counter-jet, which is not visible, we assumed an upper limit for the
surface brightness that corresponds to 0.15 mJy/beam, i.e. 1$\sigma$ noise level measured
on the image. From the brightness ratio estimated from equation~(\ref{ratio}) we
obtain $\beta$cos$\theta >$ 0.8, implying that the minimum velocity
is $\beta > 0.8$ (corresponding to a minimum bulk Lorentz factor $\Gamma$ = 2.8) and a maximum viewing angle $\theta = 36^{\circ}$. From
the minimum values derived, we can estimate a lower limit to the
Doppler factor by means of

\begin{equation}
\delta = \frac{1}{\Gamma (1 - \beta {\rm cos} \theta)}
\end{equation}

\noindent and we obtain $\delta$ $>$ 1.7. We remark that the Doppler factors estimated by means of variability and
jet/counter-jet brightness ratio are lower limits. The Doppler factor
obtained in \citet{abdo09} by modelling the SED is much larger, $\delta$ =
18. The discrepancy may be a consequence of the different emitting
regions. The radio variability is likely produced in the core component
instead of along the jet. Furthermore, the $\gamma$-ray emission can be
produced in a very compact region, as suggested by the very short variability
time-scale observed in some FSRQs \citep[e.g.][]{tavecchio10}. Such a small region is usually
self-absorbed at the typical radio frequencies (cm and mm
wavelengths). Indeed, the fit used to model the SED in \citet{abdo09}
does not take into consideration the radio data. 
On the other hand, the region responsible for the radio emission is
related to regions of the jets that are further away from the central
AGN, where the opacity is less severe.

\begin{figure*}
\centering
\includegraphics[width=11cm]{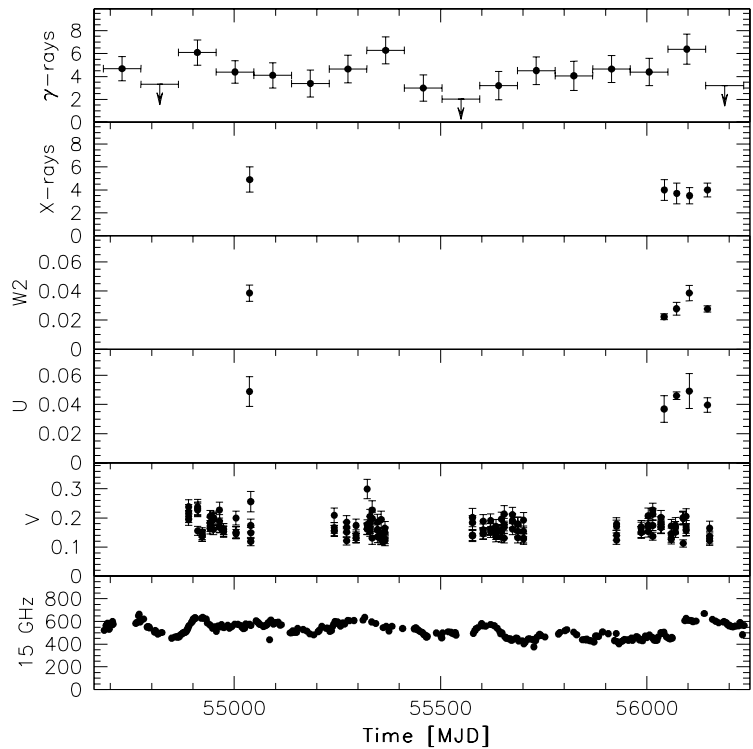}
\caption{Multifrequency light curve for PKS 1502$+$036. The period covered is 2008 August--2012 November. The data sets were collected (from top to
bottom) by {\em Fermi}-LAT ($\gamma$-rays), {\em Swift}-XRT (0.3--10 keV, in
units of 10$^{-13}$ erg cm$^{-2}$ s$^{-1}$), {\em Swift}-UVOT ($w2$ and $u$
bands, in units of mJy), CRTS ($V$ band, in units of mJy) and OVRO (15 GHz, in units of mJy).}
\label{MWL}
\end{figure*}

A black hole mass of 4$\times$10$^{6}$ M$_{\odot}$ was estimated by \citet{yuan08} for PKS 1502$+$036 on the basis of the H$\beta$ broad line. On
the other hand, \citet{abdo09} derived a black hole mass of 2$\times$10$^{7}$
M$_{\odot}$ by means of the SED modelling of the source. As in the case of SBS
0846$+$513, radio and $\gamma$-ray properties of PKS 1502$+$036 seem to show
the characteristics of a (possibly young) blazar at the low end
of the black hole mass distribution. A core-jet structure has been resolved at
15 GHz in VLBA images of PKS 1502$+$036. The main difference with respect to
SBS 0846$+$513 seems to be a fainter jet-like structure and no observed apparent superluminal motion.

In Fig.~\ref{MWL}, we compare the $\gamma$-ray light curve collected by {\em
  Fermi}-LAT during 2008 August--2012 November to the X-ray (0.3--10 keV),
UV ($m2$ filter), optical ($u$ and $V$ filters) and radio (15 GHz) light curves
collected by {\em Swift}, CRTS and OVRO. No significant flaring activity was detected in $\gamma$-rays. The lack of
significant $\gamma$-ray variability is not surprising taking into account
that only 40\% of the AGN in the second LAT AGN Catalogue (2LAC) clean sample
has shown a clear variability \citep{ackermann12b}. The continuous monitoring of the $\gamma$-ray sky provided by {\em
  Fermi}-LAT will allow us to catch, if it happens, a flaring activity from
PKS 1502$+$036. Interestingly a slight increase from radio to UV was observed at the end
of 2012 June during a period of relatively high $\gamma$-ray flux. Similarly,
the maximum flux density in $V$ band was observed when the $\gamma$-ray emission was increasing. However, the
sparse coverage in optical-to-X-rays does not allow us to obtain conclusive
evidence. A regular monitoring of this source, as well as the other
$\gamma$-ray emitting NLSy1s, from radio to $\gamma$-rays will be fundamental
for continuing to investigate the nature and the emission mechanisms of these
objects. This will be crucial for revealing differences and similarities between $\gamma$-ray NLSy1s and blazars.

\section*{Acknowledgements}

The {\em Fermi}-LAT Collaboration acknowledges generous ongoing
support from a number of agencies and institutes that have supported
both the development and the operation of the LAT as well as
scientific data analysis.  These include the National Aeronautics and
Space Administration and the Department of Energy in the United
States, the Commissariat \`a l'Energie Atomique and the Centre
National de la Recherche Scientifique / Institut National de Physique
Nucl\'eaire et de Physique des Particules in France, the Agenzia
Spaziale Italiana and the Istituto Nazionale di Fisica Nucleare in
Italy, the Ministry of Education, Culture, Sports, Science and
Technology (MEXT), High Energy Accelerator Research Organization (KEK)
and Japan Aerospace Exploration Agency (JAXA) in Japan, and the
K.~A.~Wallenberg Foundation, the Swedish Research Council and the
Swedish National Space Board in Sweden. Additional support for science
analysis during the operations phase is gratefully acknowledged from
the Istituto Nazionale di Astrofisica in Italy and the Centre National
d'\'Etudes Spatiales in France.
Part of this work was done with the contribution of the Italian Ministry of Foreign Affairs and Research for the collaboration project between Italy and Japan.
We thank the {\em Swift} team for making these observations possible, the
duty scientists, and science planners. The OVRO 40 m monitoring programme
is supported in part by NASA grants NNX08AW31G and NNX11A043G, and NSF grants AST-0808050 
and AST-1109911. The CSS survey is funded by the National Aeronautics and Space
Administration under Grant No. NNG05GF22G issued through the Science
Mission Directorate Near-Earth Objects Observations Programme.  The CRTS
survey is supported by the U.S.~National Science Foundation under
grants AST-0909182. The VLA and VLBA are operated  by the National Radio Astronomy Observatory. The NRAO is a facility of 
the National Science Foundation operated under cooperative agreement by
Associated Universities, Inc. F. D., M. O., M. G. acknowledge financial contribution from grant PRIN-INAF-2011. We thank the referee, Dr. Dirk Grupe, and F. Schinzel for helpful comments and suggestions.

\end{document}